\documentclass[12pt,a4paper]{article}%
\usepackage{amsmath,amssymb,amsfonts}
\usepackage{graphicx,epsfig}
\usepackage[margin=10pt,font=small,labelfont=bf]{caption} 
\usepackage[english]{babel}
\usepackage{slashed}

\def\be{\begin{equation}}
\def\ee{\end{equation}}
\def\ber{\begin{eqnarray}}
\def\eer{\end{eqnarray}}

\begin{document}
\begin{titlepage}
\begin{flushright}
\end{flushright}
\vskip 1.0cm
\begin{center}
{\Large \bf The NMSSM with $m_h > 115$ GeV and a moderate stop mass.} \vskip 1.5cm
{\large Leone Cavicchia}\\[1cm]
{\it Scuola Normale Superiore and INFN, Piazza dei Cavalieri 7, I-56126 Pisa, Italy} \\[5mm]
\vskip 2.0cm \abstract{We study the generic NMSSM with the coupling $\lambda S H_1 H_2$ at the limit of perturbativity, $\lambda \approx 0.7 \div 0.8$, and a moderate stop mass, $m_{\tilde{t}} \lesssim 300$ GeV. Respecting the LEP2 bounds and insisting on naturalness, we delimit the allowed region of parameter space and we study the spectrum and phenomenology of the relatively light new particles.}  
\end{center}
\end{titlepage}

\section{Introduction}
Supersymmetry provides one of the most attractive solutions to the SM hierarchy problem. The missing evidence for a light Higgs, however, introduces the need for large loop corrections driven by the stop in order to keep the Minimal Supersymmetric Standard Model (MSSM) viable. The presence of a heavy stop requires some accidental cancellations to get the correct value of the Z boson mass, reintroducing a fine-tuning in parameters at the percent level. Moreover, this residual fine-tuning weakens the argument suggesting that supersymmetry should  manifest itself around the weak scale, and thus be visible at the LHC.

As it is well known, in the most straightforward extension of the MSSM, the Next to Minimal SM (NMSSM), there is an additional quartic term in the scalar potential that may help increasing the Higgs mass without the need to invoke large loop corrections (see \cite{drees, zwirner, quiros, Pandita:1993hx, zerwas} and references therein). The NMSSM contains indeed  an extra gauge singlet chiral supermultiplet $S$, with the superpotential Yukawa interaction $\lambda S H_1 H_2$. In this framework, the mass of the SM-like Higgs, i.e. the one that couples to vector bosons before mixing, is given at tree level by:
\be
m_h^2 = M_Z^2 \cos^2 2 \beta +\lambda^2 v^2 \sin^2 2 \beta \, .
\label{hmass}
\ee
It is evident from (\ref{hmass}) that the value of the Higgs mass depends crucially on how big $\lambda$ is taken  at the weak scale. In particular, if one does not want to spoil manifest perturbative unification, using the Renormalization Group equations one finds the upper bound $\lambda \lesssim 0.6$ at low energies. Together with the complementary dependence on the angle\footnote{As usual, we define $\tan \beta =v_2/v_1$} $\beta$ of the two terms in the r.h.s. of Eq. (\ref{hmass}), this makes again difficult to push $m_h$ above the LEP2 bound without a sufficiently heavy stop.

More recently it has been  shown in \cite{pomarol, Espinosa:1998re, barbieripq} that if one assumes  the presence of extra matter at intermediate energies, filling complete $SU(5)$ supermultiplets, the RG evolution of $\lambda$ is slowed down. It is then possible to take $\lambda \approx 0.7 \div 0.8$ at the weak scale, obtaining a Higgs boson with a mass around $115 \div 125$ GeV, and a stop of moderate mass, consistently with perturbative unification. 

In \cite{barbieripq} an explicit model was presented  based on this framework, where a Peccei-Quinn symmetry in the superpotential was assumed, only weakly broken in the supersymmetry breaking terms. Other than solving the $\mu$-problem, this allows to keep under control the number of new parameters. 
This is an interesting particular example which has, however, a restricted range in parameter space where all experimental bounds are satisfied, without the need to tune some parameters at the 10 \% level.

In this work we consider the generic NMSSM with R-symmetry on the superpotential, while still keeping $\lambda \approx 0.7 \div 0.8$ at the weak scale and a moderate stop mass,  $m_{\tilde{t}} \lesssim 300$ GeV.
%In this work we study a different model, where a R-symmetry is imposed on the superpotential. 
There is in this case the additional interaction $k S^3$, that modifies the vacuum structure of the theory, opening up a wider region in parameter space. We  consider the general breaking of the R-symmetry in the soft SUSY-breaking potential, including both the trilinear terms $A_k$ and $A_{\lambda}$. Many recent works have considered the NMSSM with a spectrum of relatively light particles \cite{Dermisek2005, Chang, Dermisek2007}. We are not aware, however, of any work focussed on the generic NMSSM with $\lambda = 0.7 \div 0.8$, moderate $\tan{\beta}$ and a stop mass below 300 GeV.

The paper is organized as follows. In Sec. \ref{hsector} we present the model and we discuss the spectrum and the couplings  of the Higgs scalars and of the Higgsinos. In Sec. \ref{par} we discuss the naturalness of parameters and the bounds coming from negative searches at LEP2, and in particular their interplay in setting limits on the available parameter space. In Sec. \ref{EWPT} we present the evaluation of the contribution to the S and T parameters from  the extended Higgs sector. In Sec. \ref{LHC} we discuss some of  the experimental signatures of this model at the LHC.

\section{The Higgs-Higgsino sector}
\label{hsector}
In this Section we describe in detail the Higgs sector of the model, its parameters, the  mass spectrum and the main properties of the new particles.

\subsection{The scalar potential}
We consider the following superpotential:
\be
W = \lambda S H_1 H_2 + \frac{k}{3} S^3  \, .
\label{superpot}
\ee
The potential of Eq. (\ref{superpot}) exhibits a continuous R-symmetry, that forbids any mass term in the superpotential, providing an elegant solution to the $\mu$-problem. The $S^3$ term 
leads to relevant differences in phenomenology with respect to the PQ-symmetric limit considered in \cite{barbieripq}. The soft supersymmetry-breaking potential, that includes R-symmetry breaking terms proportional to the couplings in $W$, is given by:
\be
V_{soft}= m_1^2 |H_1|^2 + m_2^2 |H_2|^2 +m_S^2 |S|^2 + (\lambda A_{\lambda} S H_1 H_2 + \frac{1}{3} k A_k S^3 + h.c.) \, .
\ee
Assuming CP conservation, the model contains seven free parameters: $\lambda$, $k$, $m_1^2$, $m_2^2$, $m_S^2$, $A_{\lambda}$, $A_{k}$. When it exists, the CP-conserving symmetry-breaking vacuum respects the following minimization conditions:
\ber
\lambda^2 v^2 & = & \frac{m_1^2-m_2^2}{\cos(2 \beta)}+m_Z^2+\frac{2 \lambda v_s}{\sin(2 \beta)}(A_{\lambda}+k v_s) \, , \label{min1}\\
\sin(2 \beta) & = & \frac{2 \lambda v_s (A_{\lambda}+k v_s)}{m_1^2+m_2^2 +\lambda^2 v^2+2\lambda^2 v_s^2} \, , \label{min2}
\eer
\be
 2 k^2 v_s^3 + k A_k v_s^2 +v_s (\lambda^2 v^2-\lambda k v^2 \sin(2 \beta)+m_S^2) -\frac{1}{2}\lambda A_{\lambda} v^2 \sin(2\beta) = 0 \, .
\label{min3}
\ee
Using Eqs. (\ref{min1}, \ref{min2}, \ref{min3}) above, we can trade $m_1^2$ and $m_2^2$ for $v$ and $\tan \beta = v_2/v_1$.
Moreover, three parameters are fixed by requiring the mass of the SM-like Higgs to be higher than the LEP bound. 
We choose $\lambda$ and $\tan \beta$ such that they allow $m_h$ in Eq. (\ref{hmass}) above 115 GeV. The value of $\lambda$ is determined by the request of perturbativity up to the GUT scale; once $\lambda_{GUT}/4 \pi \sim 0.15 \div 0.3$ is fixed, the RG running with extra matter (three $5 + \bar{5}$ of SU(5)) gives the value at the weak scale $\lambda \approx 0.7 \div 0.8$. Moreover, the request to have $m_h > 115$ GeV in Eq. (\ref{hmass}) limits $\tan \beta$ in the interval $1.5 \lesssim \tan \beta \lesssim 2.5$, with a maximal value of $m_h$ for $\tan \beta \approx 2$. The Higgs mass does not depend directly on $k$ (before mixing). However, since the RG equations for $\lambda$ and $k$ are coupled, it can be shown that a large $k$ would reduce the maximal value of $\lambda$. In order to avoid this effect, we have to choose $k_{GUT}/4 \pi \leq 0.05$; this correspond at the weak scale to $k \lesssim 0.1$. For a detailed discussion of these aspects see \cite{barbieripq}.
Throughout the rest of the  paper, unless explicitly stated, we will consider $\lambda, k$ and $\tan \beta$ fixed at the values that maximize the Higgs mass ($\lambda=0.8, k=0.1, \tan \beta = 2$).

\begin{figure}[h]
		\begin{center}
	\includegraphics[width=6.5cm]{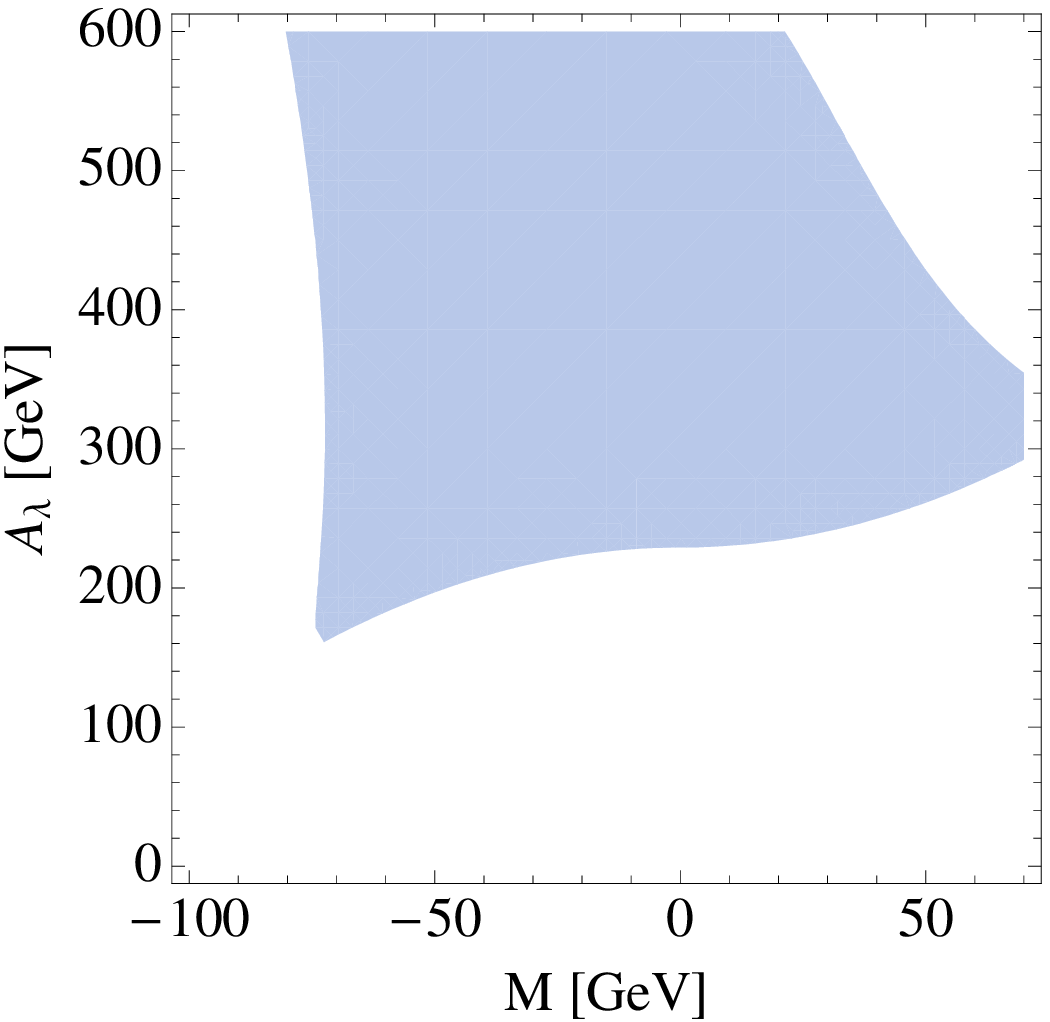}		\includegraphics[width=6.5cm]{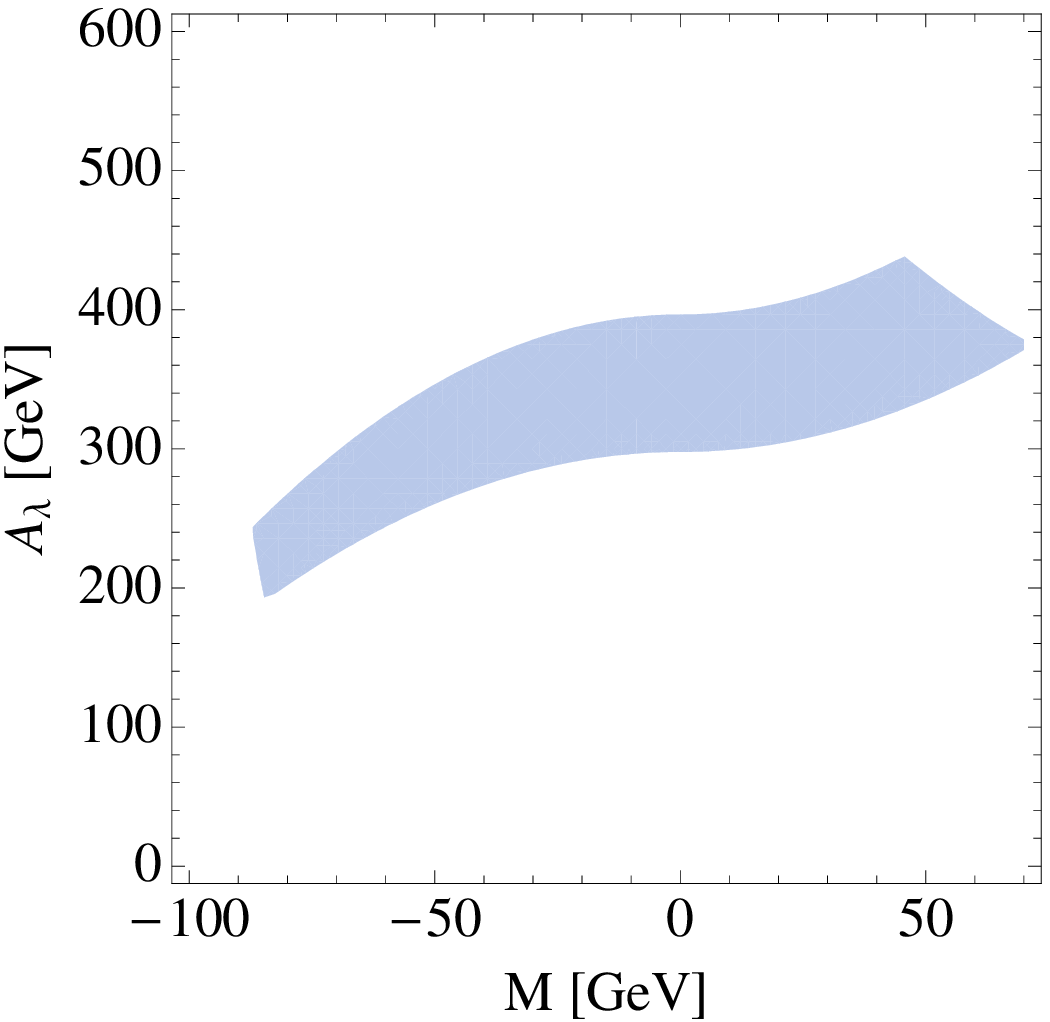}
		\end{center}
	\caption{Allowed regions (cyan shaded) in parameter space due to stability and chargino mass bounds, as a function of $A_{\lambda}$ and $M = Sgn(m_S^2) \sqrt{|m_S^2|}$, for $A_k=0$ (left) and $A_k=150$ GeV (right).}
	\label{potstab}
\end{figure}

We are thus left with only three effective free parameters in the potential: $m_S^2, A_{\lambda}, A_k$. As it can be seen in Fig. \ref{potstab}, for fixed $A_k$ only a delimited region in the $m_S^2-A_{\lambda}$ plane is allowed. This region shrinks for higher values of $A_k$. In Fig. \ref{potstab} the bound on the  right comes from global stability of the $SU(2) \times U(1)$-breaking minimum, while the ones on the left and top come from local stability\footnote{For the discussion of experimental limits on light scalars, see Sec. \ref{leplim} below.} of the same minimum. The bound on the bottom is instead a consequence of the experimental limits on the chargino mass (see Sec. \ref{leplim}).

\subsection{Spectrum}
\label{spectrum}
The Higgs multiplets $H_1, H_2$ and $S$ appearing in the potential (\ref{superpot}) contain seven bosonic physical degrees of freedom: two charged scalars, three neutral CP even and two neutral CP odd particles. Having defined
\begin{equation}
H_{1}^{0}=\frac{1}{\sqrt{2}}(h_{1}+i\pi_{1}),~H_{2}^{0}=\frac{1}{\sqrt{2}%
}(h_{2}+i\pi_{2}),~S=\frac{1}{\sqrt{2}}(s+i\pi_{s}),
\end{equation}
it is convenient to write down the scalar mass matrices in the basis where the eaten goldstone bosons decouple, given by:
\begin{equation}
	H=\cos{\beta}h_{2}-\sin{\beta}h_{1},\,h=\cos{\beta}h_{1}+\sin{\beta}%
	h_{2},\,s
	\label{sbasis}
\end{equation}
\be
P_1=\cos{\beta}\pi_{2}-\sin{\beta}\pi_{1},\,P_2=\pi_s \,.
\label{abasis}
\ee
In this basis, among the CP-even fields only $h$ has a trilinear coupling $g_{hVV}$ with the SM vector bosons.

In the following we report the mass matrices of CP odd and even states, in the basis   defined in Eqs. (\ref{sbasis}, \ref{abasis}). We keep the explicit dependence of the matrix elements upon the singlet vev $v_s$.
%since it is not possible to find a simple analytical solution of Eq. (\ref{min3}); 
Whenever it appears, $v_s$ has to be read as the numerical solution of Eq. (\ref{min3}) as a function of the free parameters $A_{\lambda}, A_k $ and $m_S^2$.
The squared mass matrix elements for the scalar degrees of freedom are, in the ($H, h, s$) basis of Eq. (\ref{sbasis}):
\ber
m^2_{11} & = & \frac{2 \lambda v_s}{\sin(2 \beta)}(A_{\lambda}+k v_s) + \sin^2(2 \beta)(m_Z^2-\lambda^2 v^2) , \nonumber\\
m^2_{12} & = & -\frac{1}{2}\sin(4 \beta)(m_Z^2-\lambda^2 v^2)  ,\nonumber\\
m^2_{13} & = & -\lambda v \cos(2 \beta)(A_{\lambda}+2 k v_s)   ,\nonumber\\
m^2_{22} & = & m_Z^2 \cos^2(2 \beta)+ \lambda^2 v^2 \sin^2(2 \beta)   ,\nonumber\\
m^2_{23} & = & 2 \lambda^2 v_s v-\lambda v \sin(2\beta)(A_{\lambda}+2 k v_s)  ,\nonumber\\
m^2_{33} & = & \frac{1}{2} \lambda A_{\lambda}\sin(2\beta) \frac{v^2}{v_s}+4 k^2 v_s^2+ k A_k v_s  .
\eer
The pseudoscalar squared mass matrix is, in the basis ($P_1$, $P_2$) of Eq. (\ref{abasis}):
\be
\mathcal{M}^2_A= \left(
\begin{array}{cc}
	\frac{2 \lambda v_s}{\sin(2 \beta)}(A_{\lambda}+k v_s) & \lambda v (A_{\lambda}-2 k v_s)\\
	\lambda v (A_{\lambda}-2 k v_s) & \frac{\sin(2\beta)}{2}v^2(\frac{\lambda}{v_s}(A_{\lambda}+ k v_s) + 3 \lambda k )-3 k A_k v_s
\end{array}
\right) .
\ee
The charged Higgs mass is given by:
\be
m^2_{H^{\pm}}= m_W^2 -\lambda^2 v^2+ \lambda (A_{\lambda} + k v_s) \frac{2 v_s}{\sin2 \beta} .
\ee

\begin{figure}[h]
		\begin{center}
	\includegraphics[width=6.5cm]{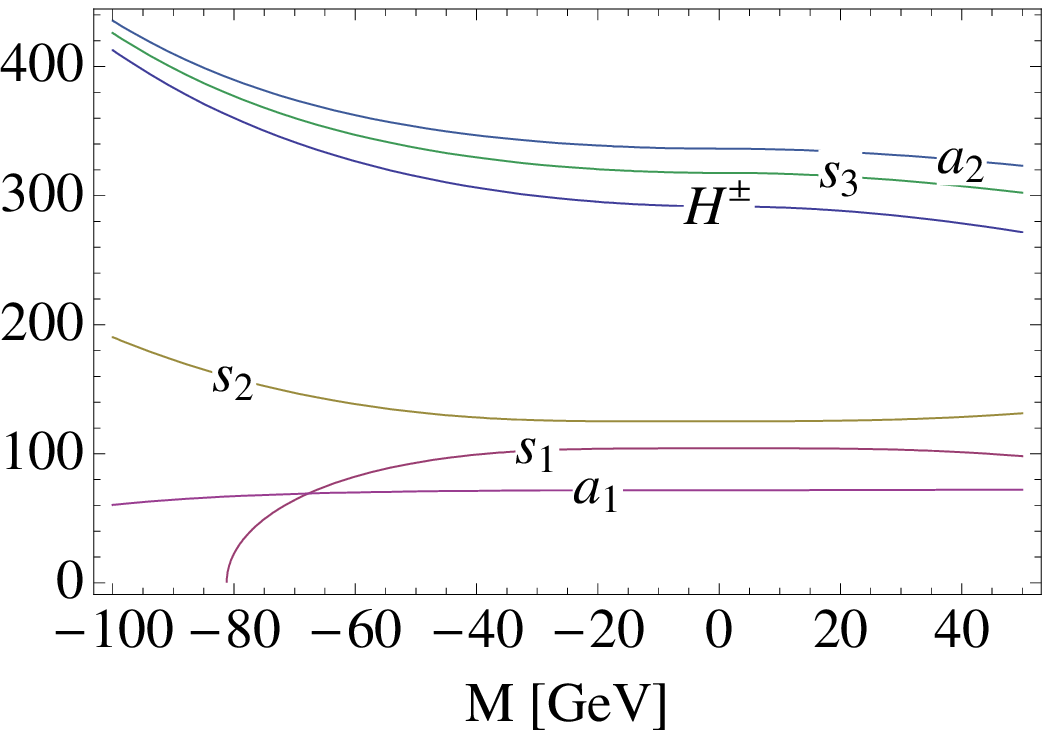}		\includegraphics[width=6.5cm]{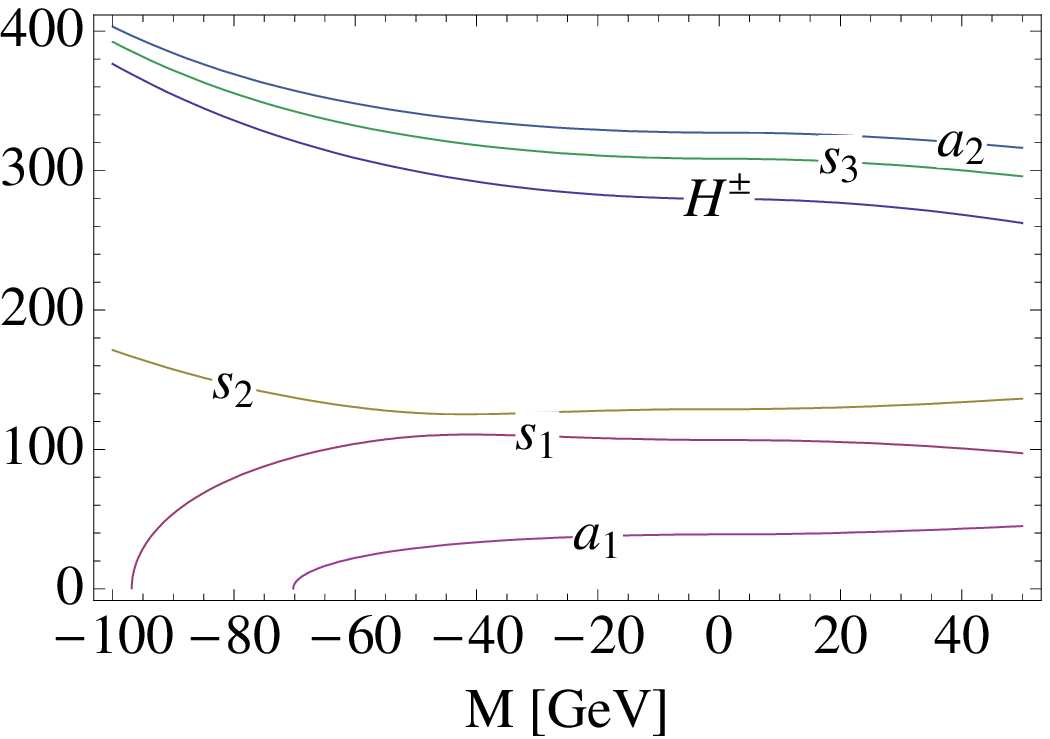}
		\end{center}
	\caption{Scalar spectrum as a function of $M = Sgn(m_S^2) \sqrt{|m_S^2|}$ for $A_k=50$ GeV (left) and $A_k=150$ GeV (right), and fixed $A_{\lambda}=300$ GeV. }
	\label{scalspect}
\end{figure}

\noindent The resulting spectrum consists of:
\begin{itemize}
	\item two relatively light scalar particles, $s_1$ and $s_2$, the first lighter, the second heavier than the LEP2 bound on the SM Higgs of about 115 GeV;
	\item a light pseudoscalar $a_1$, in the $60 \div 80$ GeV, that would be massless in presence of a PQ symmetry;
	\item three heavy, nearly degenerate states $s_3, a_2, H^{\pm}$.
\end{itemize}
 The scalar particles spectrum is shown in Fig. \ref{scalspect} as a function of $m_S^2$, for different values of $A_k$, and fixed $A_{\lambda} = 300$ GeV. The effect of a moderate loop correction ($m_{\tilde{t}}=300$ GeV) is included.
It has to be noticed that  a positive $A_k$ increases the mass of the lightest scalar state $s_1$, pushing down at the same time the light pseudoscalar $a_1$.

\subsection{The Higgsino spectrum}
\begin{figure}[h]
		\begin{center}
	\includegraphics[width=9cm]{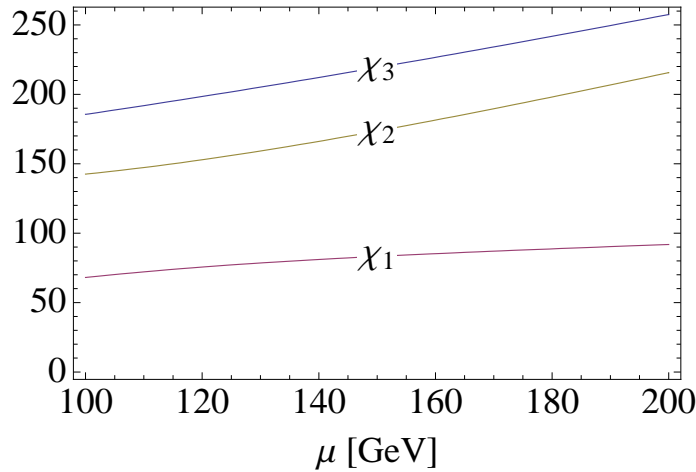}		
		\end{center}
	\caption{Higgsino spectrum as a function of the chargino mass $\mu$.}
	\label{fermspect}
\end{figure}

In the Higgsino sector, we  assume that the gaugino mass parameters $M_1$ and $M_2$ are large, so that gaugino-Higgsino mixing is negligible and the chargino mass can be kept  above the LEP2 limit. 
In the basis
\be
 N_{1} = \frac{1}{\sqrt{2}}(\tilde{H_{1}}-\tilde{H_{2}}),
\quad
  N_{2} = \frac{1}{\sqrt{2}}(\tilde{H_{1}}+\tilde{H_{2}}),
\quad
  N_{3} = \tilde{S},
\ee
the Higgsino mass matrix is given by:
\be
\mathcal{M}_{\chi}= \left(
\begin{array}{ccc}
 \mu & 0 & \frac{\lambda v}{\sqrt{2}} (\cos\beta-\sin\beta)\\
0 & -\mu & -\frac{\lambda v}{\sqrt{2}} (\cos\beta+\sin\beta)\\
\frac{\lambda v}{\sqrt{2}} (\cos\beta-\sin\beta)& -\frac{\lambda v}{\sqrt{2}} (\cos\beta+\sin\beta) & 2 k v_s 
\end{array}
\right) .
\label{neutrmass}
\ee
The mass eigenstates $\chi_i$ are found applying  a rotation that diagonalizes the $\mathcal{M}_{\chi}$ matrix, such that:
\be
\chi_i = V_{ij} N_j \quad V V^T=1 \quad V \mathcal{M}_{\chi}V^T=\mathcal{M}_{\chi}^D .
\label{neutrdiag}
\ee
The resulting mass spectrum consists of a light, mostly singlino, state and two heavier states, separated in mass by about $40 \div 50$ GeV (see Fig. \ref{fermspect}).

\noindent The chargino mass is given (in the heavy gaugino approximation) by
\be
m_{\chi^{\pm}}= \mu.
\label{chargmass}
\ee
The effective $\mu$ term appearing in (\ref{neutrmass}) and (\ref{chargmass}) is given by
\be
\mu = \lambda v_s,
\ee
and thus no new parameters are introduced in the Higgsino sector.

\subsection{Couplings and decays}
\label{coupl}
\begin{figure}[h]
		\begin{center}
	\includegraphics[width=6.5cm]{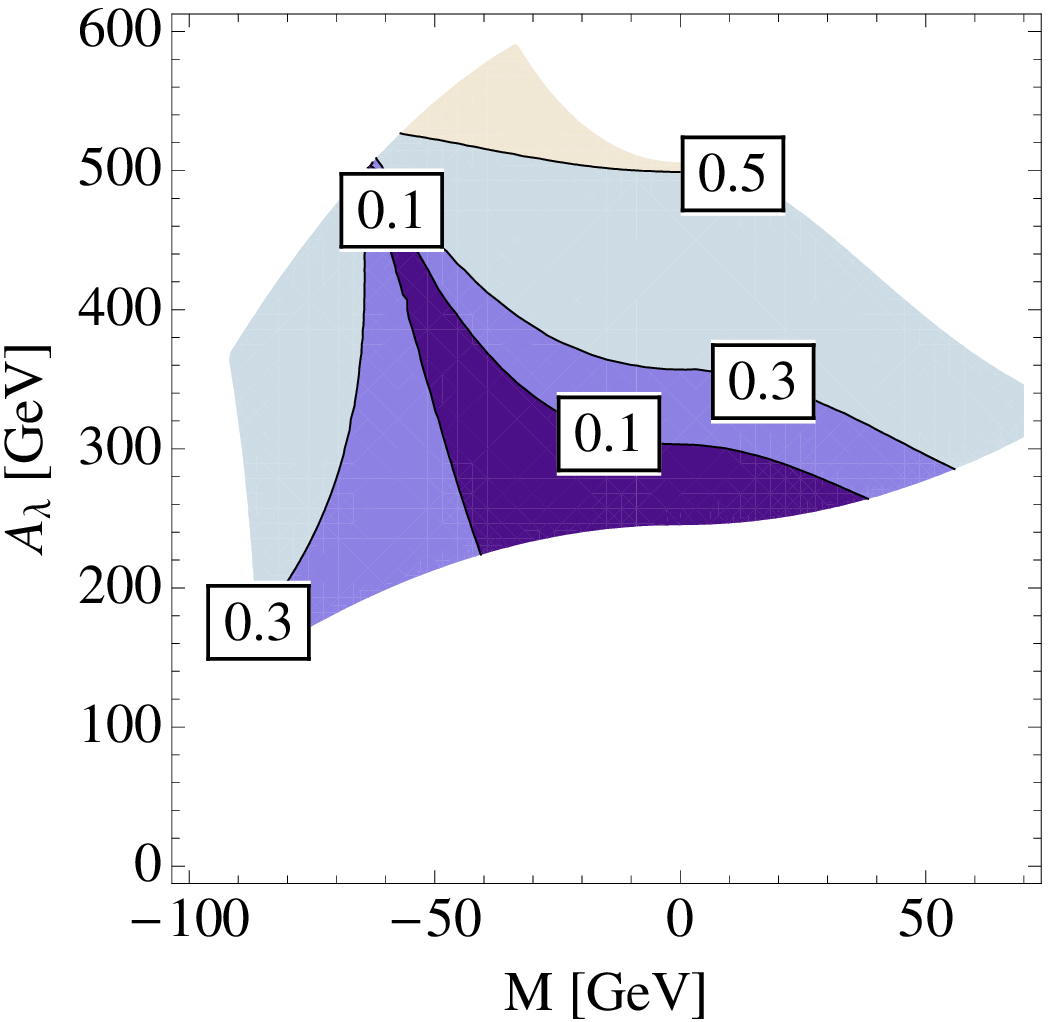}				\includegraphics[width=6.5cm]{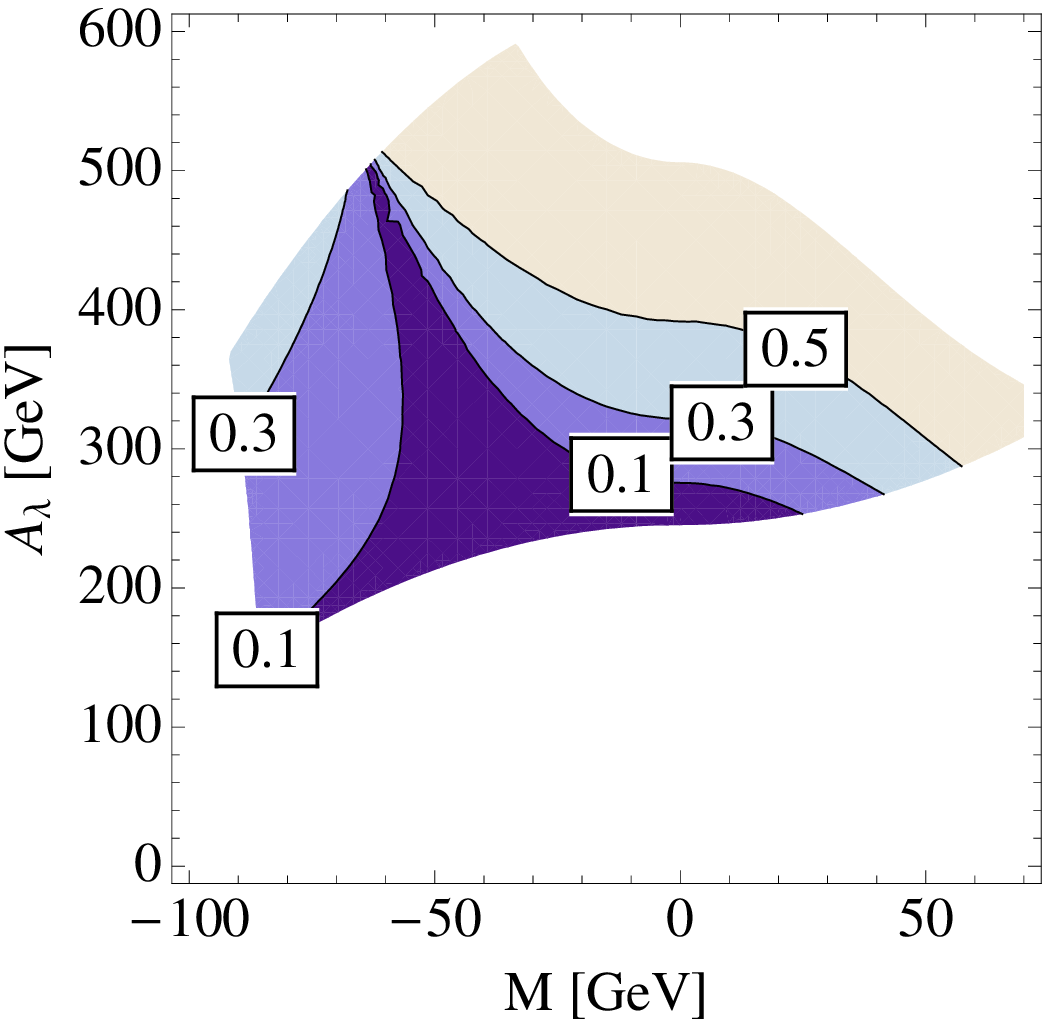}		
		\end{center}
	\caption{Squared normalized couplings of $s_1$. Coupling to vectors pairs $\xi^2_{s_1VV}$ (left) and top pairs $\xi^2_{s_1t\bar{t}}$ (right), plotted  as a function of $M = Sgn(m_S^2) \sqrt{|m_S^2|}$ and $A_{\lambda}$, for fixed $A_k=100$ GeV.}
	\label{h1vv}
\end{figure}

In this section we discuss the properties of the Higgs sector particles relevant for phenomenology, such as production couplings and branching ratios.
A few characteristic properties have to be noticed, crucial to  the collider phenomenology. The lightest scalar, $s_1$, acquires a trilinear coupling to vectors $g_{s_1VV}$ only through the mixing with the SM-like Higgs; the resulting suppression in production rates is shown in Fig. \ref{h1vv}, where the squared normalized coupling $\xi^2= g^2_{SUSY}/g^2_{SM}$ is plotted . Once produced, this particle will decay in light pseudoscalar pairs $a_1 a_1$, whenever the channel is kinematically allowed. Otherwise, it will decay predominantly in $b\bar{b}$  pairs, with the remaining part decaying in $\tau^+ \tau^-$ with a relative weight given by 
\be
\frac{BR(\tau^+ \tau^-)}{BR(b\bar{b}) }\sim \left(\frac{m_{\tau}}{m_b}\right)^2.
\label{btaubr}
\ee
The light pseudoscalars $a_1$ will in turn decay mostly in $b\bar{b}$ with the same BR of Eq. (\ref{btaubr}).

% \begin{figure}[h]
%		\begin{center}
%	\includegraphics[width=6.5cm]{brh1aa75}				\includegraphics[width=6.5cm]{brh1aa125}	
%			\end{center}
%	\caption{Branching ratio of $s_1$ in $a_1$ couples as a function of $M$ (see \ref{mplot}) and $A_{\lambda}$, for $A_k=75$ GeV (left) and $A_k=125$ GeV (right).}
%	\label{h1aa}
% \end{figure}

The scalar $s_2$ has nearly full strength couplings to SM particles. Its production rates are therefore close to the SM ones. Moreover,  the decay channel of $s_2$ into neutralinos is always closed, due to the enhancing effect of the $k$ term on the mass of the lightest neutralino. On the other hand, in wide regions of parameter space the decay into light pseudoscalar pairs $a_1 a_1$ is dominant. Except for the region $m_S^2 \lesssim -(80 \mbox{ GeV})^2$ where the channel $s_2\rightarrow s_1 s_1$ becomes also kinematically accessible, the fraction $1-BR(s_2\rightarrow a_1 a_1)$ has SM-like decays.

\begin{figure}[h]
		\begin{center}
	\includegraphics[width=6.5cm]{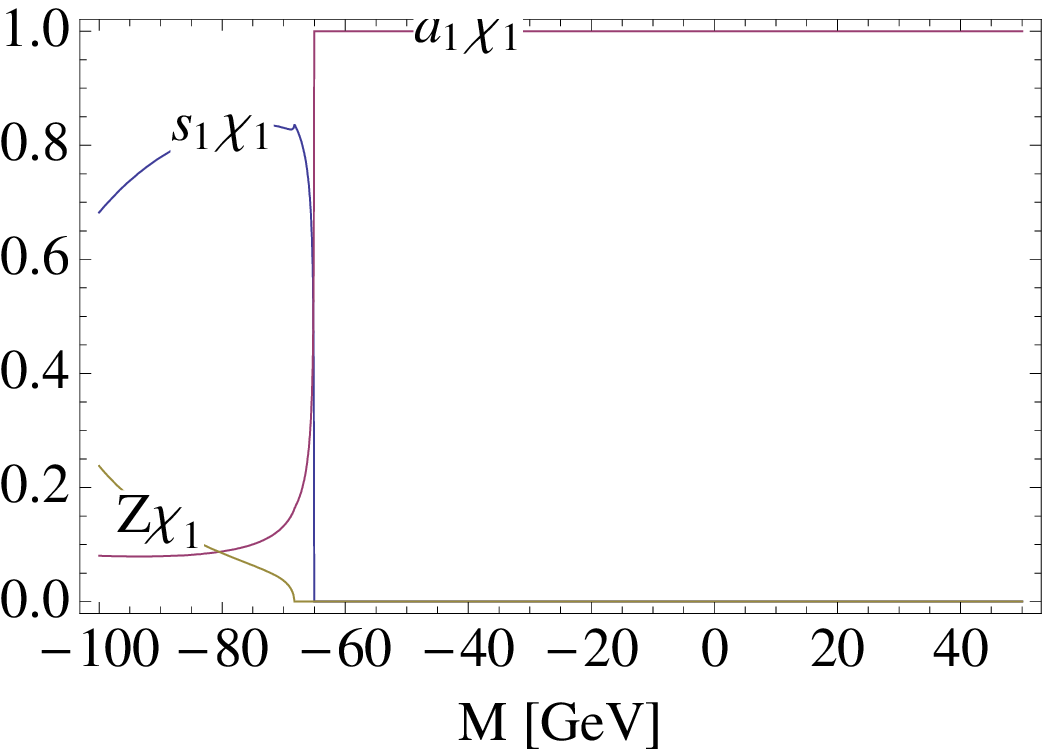}				\includegraphics[width=6.5cm]{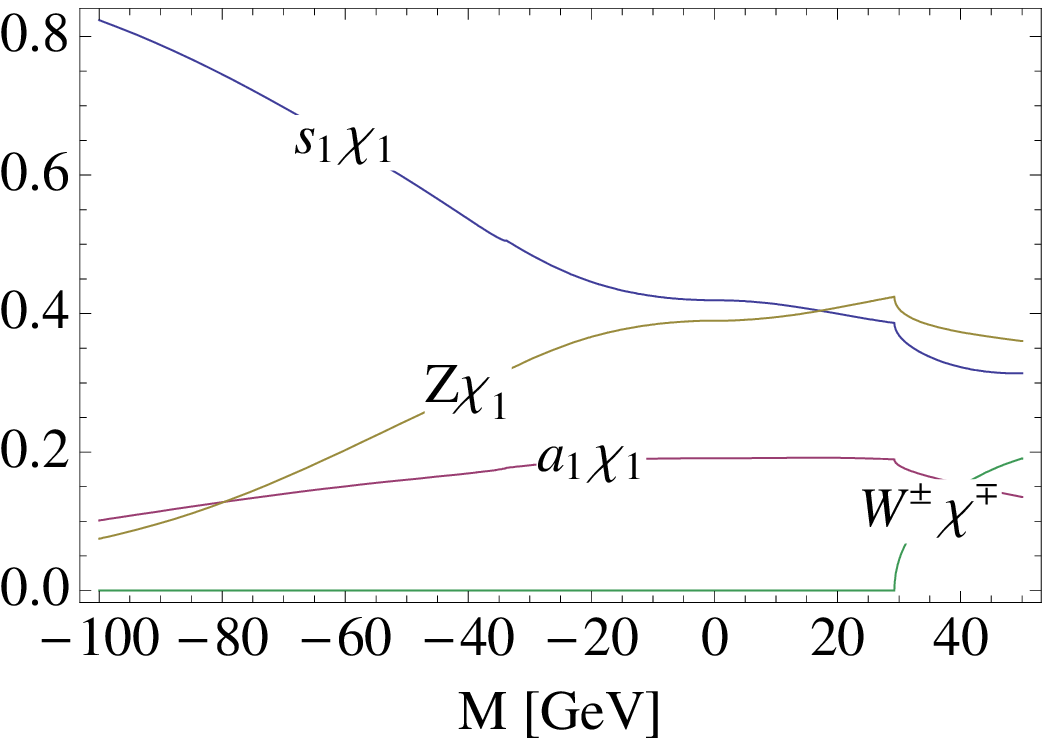}	
			\end{center}
	\caption{Typical branching ratios of heavy neutralinos $\chi_2$ (left) and $\chi_3$ (right) as a function of $M = Sgn(m_S^2) \sqrt{|m_S^2|}$.  ($A_{\lambda}=300 \mbox{ GeV}, A_k=100$ GeV).}
	\label{xdec}
\end{figure}

Another interesting property is that the two heavy neutralinos will decay with a sizable branching fraction in scalar particles (see Fig. \ref{xdec}). This decays go through the singlino component of $\chi_i$ .

\begin{figure}[h]
		\begin{center}
					\includegraphics[width=8cm]{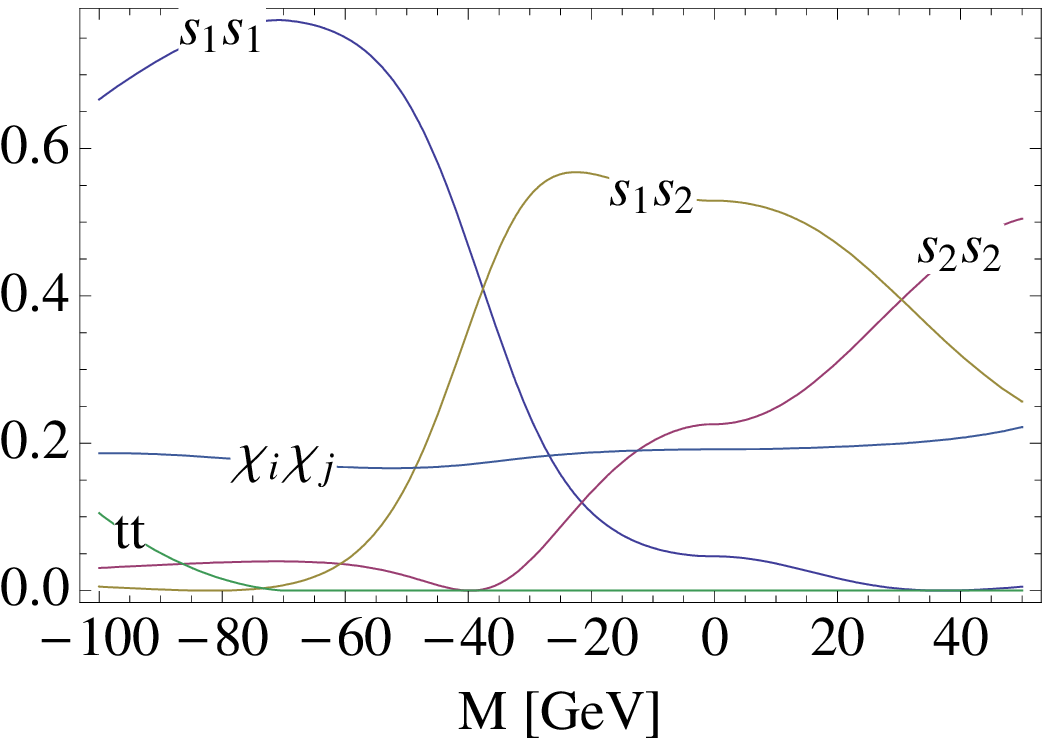}	
			\end{center}
	\caption{Typical branching ratios of the heavy scalar  $s_3$  as a function of $M = Sgn(m_S^2) \sqrt{|m_S^2|}$. ($A_{\lambda}=300 \mbox{ GeV}, A_k=100$ GeV).}
	\label{hdec}
\end{figure}

The branching fractions of the remaining heavy states are summarized in Figs. \ref{xdec} and \ref{hdec}.

%In this kind of scenarios a  light gravitino is present, that takes the role of the LSP,
%while the lightest neutralino is  the NLSP. A very critic property for phenomenology is thus the lifetime of the $\chi_0^1$. If this particle is long lived, it will %decay to gravitino well outside the detector and can thus be considered from the phenomenology point of view as if it was the LSP itself; otherwise, if it decays 
% inside the detector, the phenomenology will appear quite different.
% This aspects will be further discussed in Sec. \ref{LHC}.

\section{Parameter space: naturalness and LEP constraints}
\label{par}
In this section we proceed further with the analysis of the parameter space, in order to check whether regions exist that are not yet excluded by the LEP2 searches, consistently with a moderate fine-tuning on parameters.

\subsection{Naturalness}
\label{ft}
Since one of the motivations to consider this model is the attempt to ameliorate the naturalness problem of the MSSM, it is important to look carefully at all possible sources of fine-tuning and check if regions in parameter space exist where the residual fine-tuning is moderate \footnote{A similar analysis was performed in \cite{toro}; however no strict bounds where set on models with large $\lambda$.}.

\begin{figure}[h]
			\begin{center}
		\includegraphics[width=12cm]{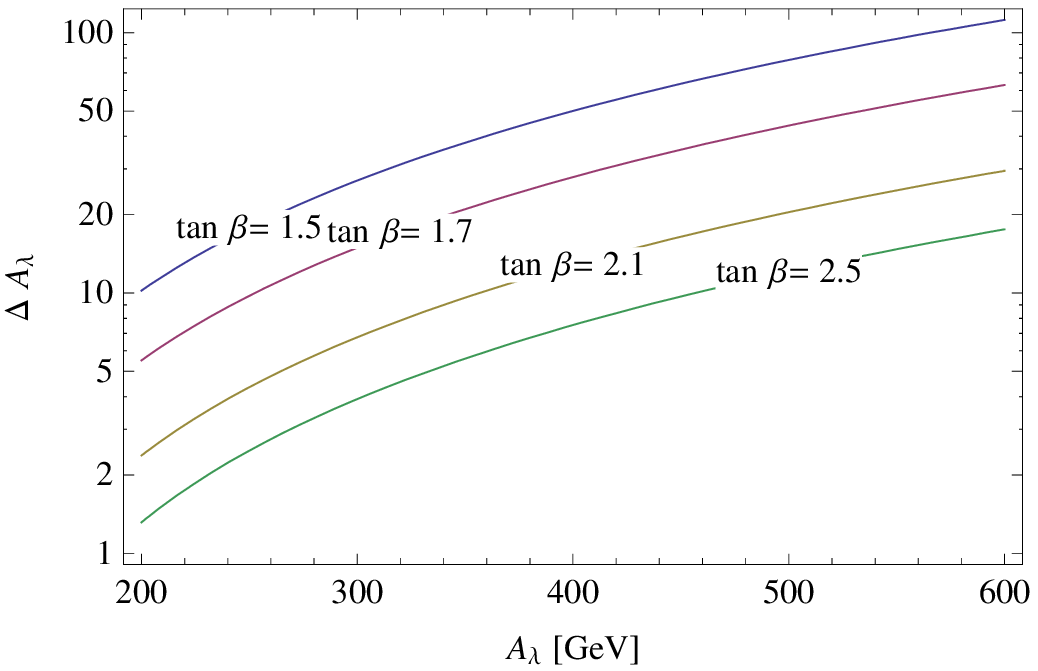}
			\end{center}
		\caption{$\Delta_{A_{\lambda}}$ as a function of $A_{\lambda}$ for $A_k = 100$ GeV  and $m_S^2=-(60 \mbox{ GeV })^2$. This quantity is almost independent on $m^2_S$ and $A_k$.}
		\label{fta}
	\end{figure}

\begin{figure}[h]
			\begin{center}
		\includegraphics[width=9cm]{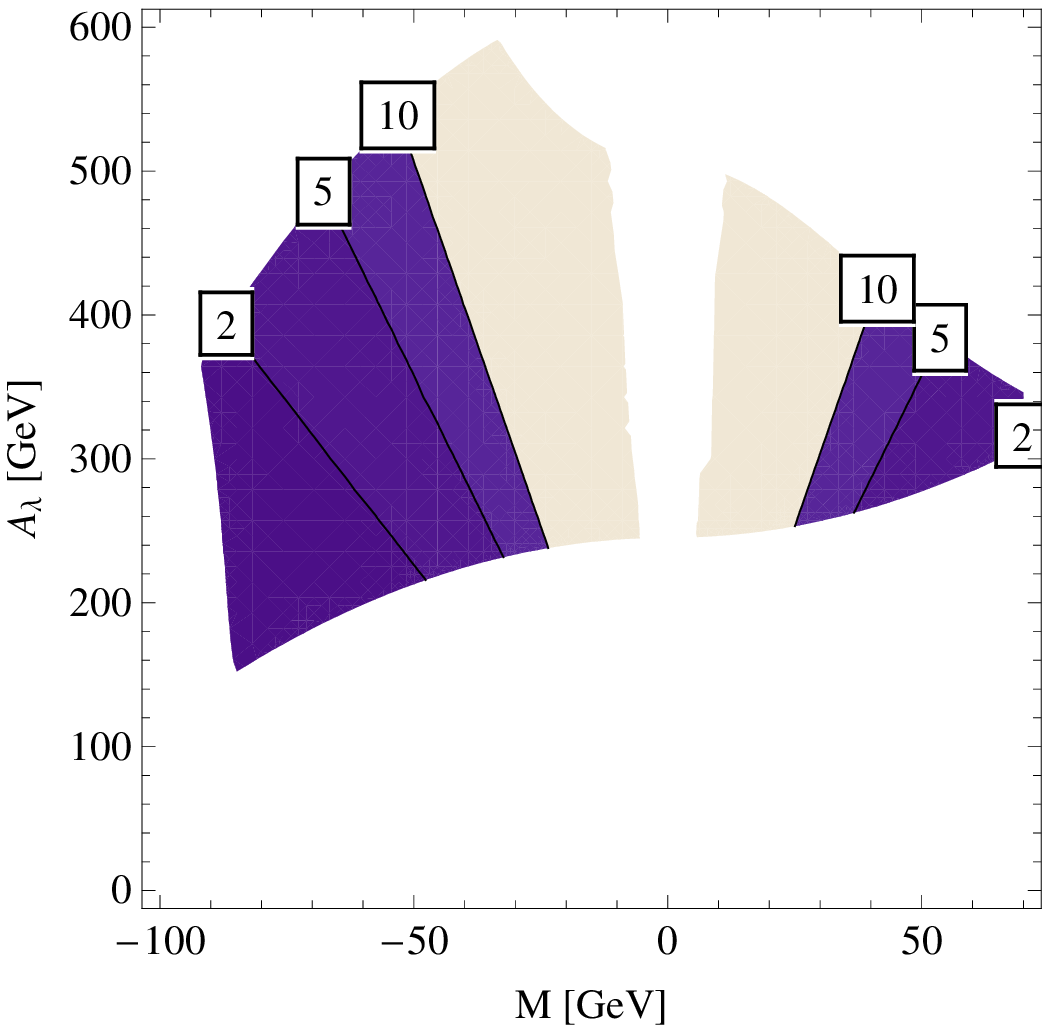}
			\end{center}
		\caption{$\delta m_S^2/m_S^2$ as a function of $M = Sgn(m_S^2) \sqrt{|m_S^2|}$  and $A_{\lambda}$ ($A_k = 100$ GeV).}
		\label{ms}
	\end{figure}

The first thing to check is the dependence of the weak scale $v$ on the various parameters $p_i$. This can be estimated by evaluating the logarithmic derivative of $v^2$ in Eq. (\ref{min1})
\be
\Delta_{p_i}=\left|\frac{\partial \log v^2}{\partial \log p^2_i}\right|,
\ee 
taking into account the variation of $\beta$ from Eq. (\ref{min2}). The only relevant such dependence is found to be the one on $A_{\lambda}$.
As it is shown in Fig. \ref{fta}, the request of a fine-tuning lower than $\sim 10 \%$  disfavor  values of $\tan \beta$ smaller than 2 and $A_{\lambda}$ bigger than $400 \div 500$ GeV.
	 The other sizable source of fine-tuning comes from the one loop contribution to the running of $m_S^2$ driven by $A_{\lambda}$. 
	In the example of a PQ-symmetric potential studied in \cite{barbieripq}, it turned out that this was the main source of fine-tuning; the ratio $\delta m_S^2/m_S^2$ in fact was always bigger than 10, for a low mediation scale of $\Lambda_{mess} \sim 100 $ TeV. In the case of R-symmetric potential we are considering, there are instead sizable regions in the $m_S^2 - A_{\lambda}$ plane, corresponding to negative $m_S^2$ values, where this ratio takes  smaller values.  Fig. \ref{ms}  shows the ratio $\delta m_S^2/m_S^2$ for $\Lambda_{mess} \sim 100 $ TeV.

\subsection{LEP limits}
\label{leplim}
The negative searches for supersymmetric particles performed by the LEP2 experiments placed several lower bounds on the sparticle masses. Since the spectrum of our model contains some light particles, it has to be checked whether they are not already excluded by LEP, and what further restrictions on parameter space this implies. 

In the Higgsino sector both the processes  $e^+ e^- \rightarrow Z^* \rightarrow \chi_1 \chi_2$ and  $e^+ e^- \rightarrow Z^* \rightarrow \chi^+ \chi^-$ have been searched for at LEP \cite{neutralino1, neutralino2, neutralino3}. Given the Higgsino spectrum of our model (Fig. \ref{fermspect}), direct $\chi_1 \chi_2$ production is not kinematically accessible. The limit on the chargino mass is $m_{\chi^{\pm}} > 103 $ GeV. Since $m_{\chi^{\pm}}=\mu$ under the assumption of heavy gauginos, we obtain the bound in the bottom of $m_S^2 - A_{\lambda}$ plane already shown in Fig. \ref{potstab}.

\begin{figure}[h]
		\begin{center}
	\includegraphics[width=6.5cm]{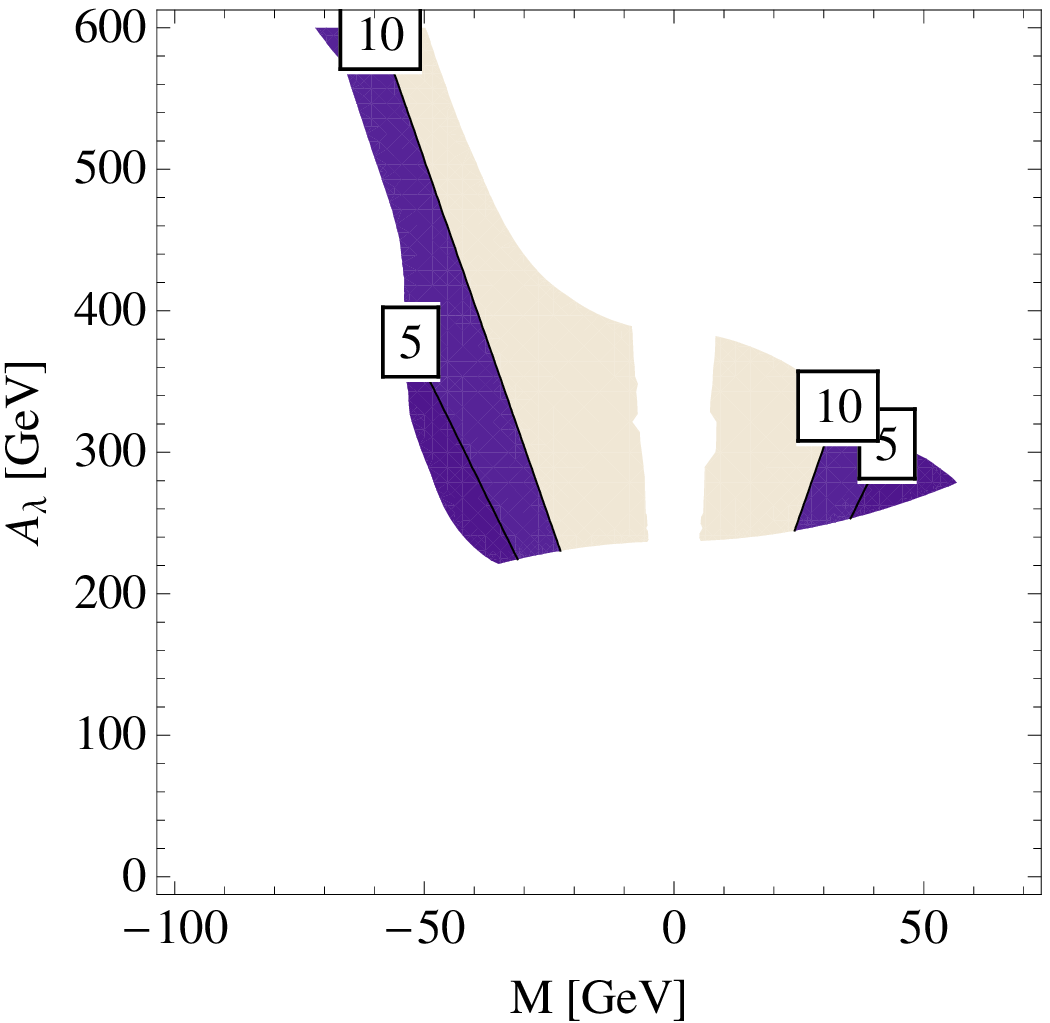}				\includegraphics[width=6.5cm]{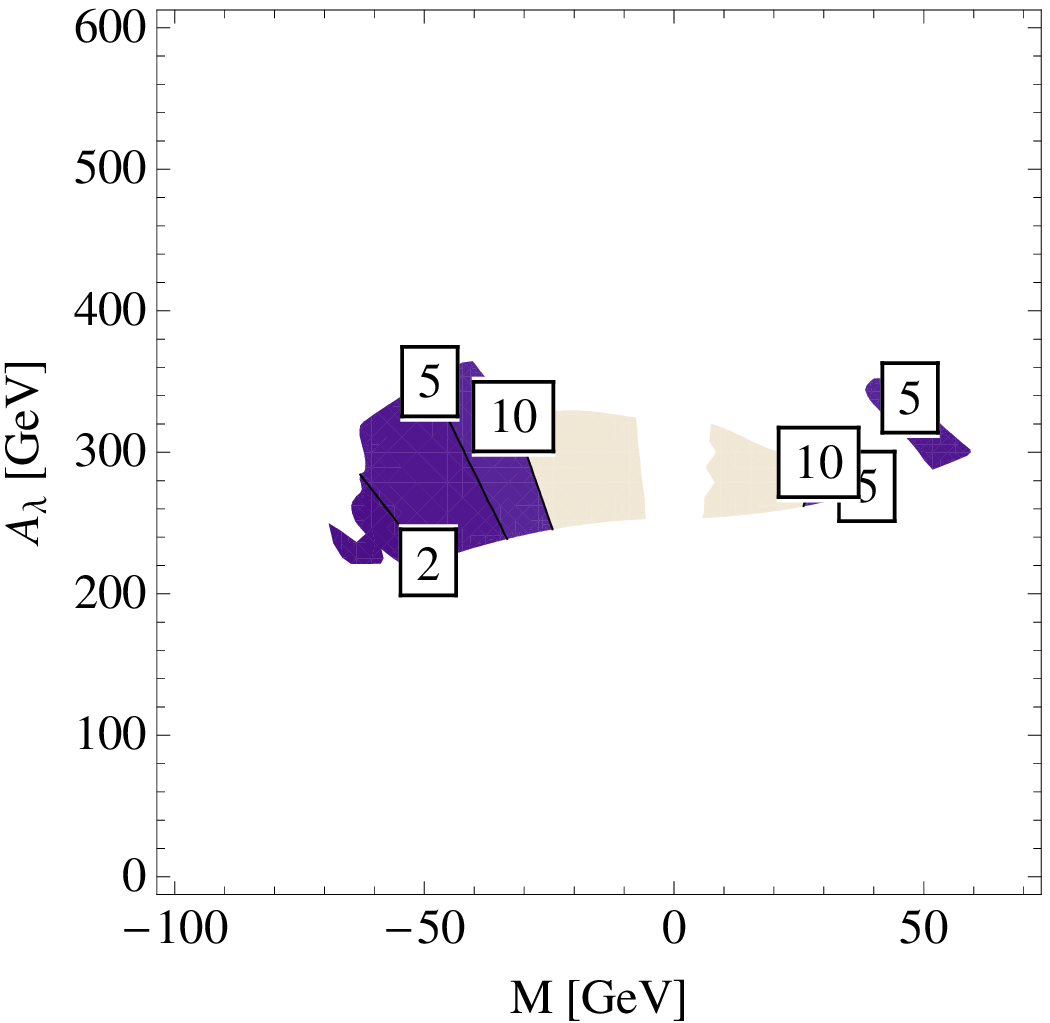}	
			\end{center}
	\caption{Contours of the ratio $\delta m_S^2/m_S^2$ in the region not excluded by LEP, as a function of $M = Sgn(m_S^2) \sqrt{|m_S^2|}$  and $A_{\lambda}$, for $A_k=50$ GeV (left) and $A_k=150$ GeV (right).}
	\label{lepms}
\end{figure}

Another process that sets stringent restrictions on the parameters is  $e^+ e^- \rightarrow Zs_1$ \cite{higgslep}. As we discussed in sec. \ref{coupl}, once produced the light scalars $s_1$ decay either directly in $b\bar{b}$ pairs or in $s_1 \rightarrow a_1 a_1 \rightarrow b\bar{b}b\bar{b}$, depending on available phase space. It is interesting to confront the regions in the  $m_S^2 - A_{\lambda}$ plane excluded by LEP with the ones compatible with a moderate fine-tuning on $m_S^2$. In fact the regions with big negative $m_S^2$ where $\delta m_S^2/m_S^2$ is low, correspond to the $s_1$ mass approaching the stability boundary $m_{s_1}^2 = 0$. One would thus naively expect that the same regions are already ruled out by LEP. 
%However, , for moderate and positive values of $A_k$ the mass of the light scalar is raised; 
It turns out, however, that for $A_k$ in the range $A_k \approx 50 \div 150$ sizable region survive with moderate fine-tuning, as it is shown in Fig. \ref{lepms}. It has to be noticed that a low value of $A_k$ is not unnatural since it is the breaking parameter associated to $k$, that as we already discussed has to be taken small by itself.

Finally, the process $e^+ e^- \rightarrow Z^* \rightarrow a_1 s_1$ is also constrained by LEP,  but, due to the extreme weakness of the $g_{Zs_1a_1}$ coupling, it does not set further bounds on parameter space.

\section{Electroweak Precision Tests}
\label{EWPT}
Besides the direct experimental constraints discussed in Sec. \ref{leplim}, we should also consider the contributions that arise from the extended Higgs-Higgsino sector to the electroweak precision parameters, S and T.  The presence of a sizable Yukawa coupling $\lambda \approx 0.8$ may in fact induce significant effects.

\noindent The contributions to S and T from scalar particles are given by:

\begin{eqnarray}
	T_{\mbox{scal}} &=& \sum_{i=1}^3\xi_{VVs_i}^2 3 (A(m_{s_i}, m_Z)-A(m_{s_i}, m_W)) +\sum_{i=1}^3\xi_{WH^{\pm}s_i}^2  A(m_{H^\pm}, m_{s_i}) \nonumber\\
	&+& \sum_{j=i}^2\xi_{WH^{\pm}a_j}^2  A(m_{H^\pm}, m_{s_i})-\sum_{i=1}^3\sum_{j=1}^2\xi_{Zs_ia_j}^2  A(m_{a_j}, m_{s_i}) ,
	\label{thiggs}
\end{eqnarray}
\begin{eqnarray}
	S_{\mbox{scal}}&=& \sum_{i=1}^3\xi_{VVs_i}^2 3 (F(m_{s_i}, m_Z)+m_Z^2 G(m_{s_i}, m_W)) \nonumber\\ &+& \sum_{i=1}^3 \sum_{j=1}^2\xi_{Zs_ia_j}^2  F(m_{a_j}, m_{s_i})
	- F(m_{H^\pm},m_{H^\pm})  ,\label{sthiggs}
\end{eqnarray}
\noindent where $s_i$ ($a_j$)   are the scalar (pseudoscalar) mass eigenstates,  the \emph{reduced} couplings $\xi$ are  defined in Tab. \ref{redcoup}, while the complete expressions for the loop functions $A, F, G$ can be found in App. \ref{stloop}.

\begin{table}
	\begin{center}
\begin{tabular}{|c|c|}
	\hline
	$\xi_{VVs_i}= U_{i2}$ & $ \xi_{WH^{\pm}} = U_{i1} $ \\ 
	\hline
	$\xi_{Zs_ia_1}= U_{i1} \cos \alpha $ & $\xi_{Zs_ia_2}= U_{i1} \sin \alpha$\\
	\hline
	$\xi_{WH^{\pm}a_1} = \cos \alpha $ & $\xi_{WH^{\pm}a_2} = \sin \alpha $ \\
	\hline
\end{tabular}
\end{center}
\caption{Reduced scalar-vectors couplings ($\xi=g_{SUSY}/g_{SM}$).}
\label{redcoup}
\end{table}

\noindent The coefficients $U_{ij}$ in Tab. \ref{redcoup} are the elements of the matrix that rotates from the basis $(H, h, s)$ of Eq. (\ref{sbasis}),
where only $h$ couples to vectors, to the mass eigenstates basis. The rotation angle $\alpha$ is defined in an analogous way as:
\begin{align}
a_1  &  =-\sin\alpha\pi_{s}+\cos\alpha(\cos\beta\pi_{2}-\sin\beta\pi
_{1}),\label{Acomp}\\
a_2  &  =\cos\alpha\pi_{s}+\sin\alpha(\cos\beta\pi_{2}-\sin\beta\pi_{1}).
\end{align}

\noindent The Higgsino contributions to S and T are given by
\begin{align}
  T_{\text{Higgsinos}} =
    & \sum\nolimits_{i=1}^{3} (V^T_{1i})^{2} \tilde{A}(\mu,m_{i})
      + (V^T_{2i})^{2} \tilde{A}(\mu,-m_{i})
\nonumber\\
    & - \frac{1}{2} \sum\nolimits_{i,j=1}^{3} (V^T_{1i}V^T_{2j}+V^T_{1i}V^T_{2j})^{2}
        \tilde{A}(m_{i},-m_{j}), \label{thiggsino}
\\
  S_{\text{Higgsinos}} =
    & \frac{1}{2} \sum\nolimits_{i,j=1}^{3} (V^T_{1i}V^T_{2j}+V^T_{1i}V^T_{2j})^{2}
        \tilde{F}(m_{i},-m_{j}) - \tilde{F}(\mu,\mu),\label{sthiggsino}
\end{align}
where $\mu$ is the chargino mass, the Higgsino/neutralino rotation matrix $V$ has been defined in (\ref{neutrdiag}), and the complete expressions for the loop functions $\tilde{A}, \tilde{F}$ can be found in App. \ref{stloop}.
%\be
%V= \mathcal{N}^T .
%\ee

\begin{figure}[h]
		\begin{center}
	\includegraphics[width=12cm]{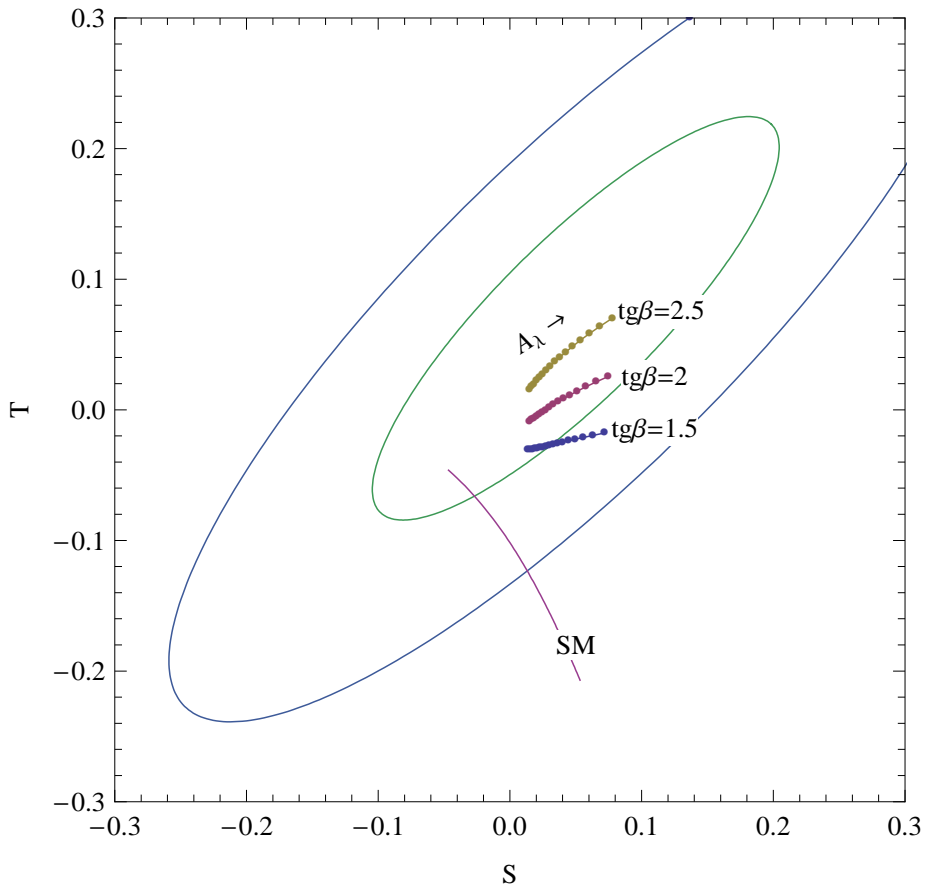}
		\end{center}
	\caption{S and T values as a function of $A_{\lambda}$ (in the range 200 GeV - 600 GeV) and $\tan \beta$, compared to the SM result and experimental limit ($A_k = 100$ GeV, $m_S^2=-(60 \mbox{ GeV})^2$).}
	\label{ewpt}
\end{figure}

In order to compare our results with experimental constraints, we want to place the contributions  from the model in the S-T plane.
This is done in two steps:
\begin{enumerate}
	\item The contributions of a SM Higgs 
	\begin{align}
	  T_{\text{Higgs}}(m_{h}) &= -3\left[ A(m_{h},m_{W})-A(m_{h},m_{Z})\right], 
\\
	  S_{\text{Higgs}}(m_{h}) &= F(m_{h},m_{Z}) + m_{Z}^2 G(m_{h},m_{Z})
	\end{align}
	are subtracted from the SM S-T values of \cite{LEPEWWG} , for a reference value of the Higgs mass;
	\item The contributions (\ref{thiggs}, \ref{sthiggs}, \ref{thiggsino}, \ref{sthiggsino}) from our model are then added.
\end{enumerate}
Fig. \ref{ewpt} shows the result as a function of $A_{\lambda}$ and $\tan \beta$, for $\lambda=0.8$.
As can be clearly seen the S and T values of our model are fully compatible with the experimental contours, for all relevant values of the free parameters (the dependence of S and T on $A_k$ and $m_S^2$ is indeed rather mild).

\section{LHC Phenomenology}
\label{LHC}
In this section we discuss some of the experimental signatures that should be looked for at the LHC, if this model is realized in Nature. We will mainly focus on the most promising channels for SUSY discovery at low luminosity.

A preliminary remark is the following. As indicated in Sec. \ref{ft}, naturalness considerations prefer (although only logarithmically) a relatively low mediation scale $\Lambda_{mess}$, and in turn a low scale of supersymmetry breaking $\sqrt{F}$. In this case a  light gravitino is present, that takes the role of the LSP,
while the lightest neutralino is  the NLSP. A crucial property for phenomenology is thus the lifetime of the $\chi_1$. If this particle is long lived, it will decay to gravitino well outside the detector and can thus be considered in the experimental signatures as the effective LSP; otherwise, if it decays 
 inside the detector, the phenomenology will appear quite different.

For an Higgsino-like neutralino, as we assumed throughout this paper, the only possible decays are $\chi_1 \rightarrow \phi \tilde{G}$, where $\tilde{G}$ is the gravitino and $\phi$ a generic scalar. In particular, given the scalar spectrum of our model, the only kinematically allowed 2-body decay is the one involving a light pseudoscalar $a_1$. The decay width is given by
\be
\Gamma =\frac{g^2_{\tilde{G} \chi_1 a_1}}{32 \pi} \frac{m^5_{\chi_1}}{ (\sqrt{F})^4}\left(1-\frac{m_{a_1}^2}{m^2_{\chi_1}}\right)^4 ,
\ee
with the coupling $g^2_{\tilde{G} \chi_1 a_1}$ given by\footnote{Mixings are defined in Eqs. (\ref{Acomp}, \ref{neutrdiag}).}
\be
g^2_{\tilde{G} \chi_1 a_1}=|\cos \alpha (V_{11} \sin \beta - V_{12}  \cos \beta) + V_{13} \sin \alpha|^2.
\ee
The neutralino decay length can be expressed as
\be
L = \frac{1.97 \times 10^{-2}}{g^2_{\tilde{G} \chi_1 a_1}} \left( \frac{E^2}{m^2_{\chi_1}}-1\right)^{1/2} \left(\frac{m_{\chi_1}}{100 \mbox{ GeV}}\right)^{-5} \left(\frac{\sqrt{F}}{100 \mbox{ TeV}}\right)^{4} \left(1-\frac{m_{a_1}^2}{m^2_{\chi_1}}\right)^{-4} \hspace{-.2cm}\mbox{cm},
\label{chidecay}
\ee
where $E$ is the neutralino energy in the lab frame.
Depending on the SUSY breaking scale $\sqrt{ F }$, Eq. (\ref{chidecay})  leads to a decay inside or outside the detector.
The critical value for $\sqrt{ F }$ is a few hundreds TeV.

In the following we will refer to the case where the $\chi_1$ lifetime is long enough such that it decays outside the detector and can be considered like the effective LSP. Otherwise, the experimental signatures will include four additional $b$-jets (originated in two displaced secondary vertexes) coming from the decay chain $\chi_1 \rightarrow a_1 \tilde{G} \rightarrow b \bar{b} \tilde{G}$. These additional jets might allow for a more efficient signal/background discrimination. However the discussion of such complicated topologies is beyond the scope of this work.

The interesting signatures for SUSY discovery depend on the full spectrum of supersymmetric particles, including gluino, squarks and sleptons. 
% We will once again stick to the naturalness criterion, and 
In order to ease possible problems in the flavor sector, and to simplify the discussion, we will assume that all those particles are heavy (in the 1-2 TeV range), except for the ones that cannot be taken heavy without introducing additional fine-tuning \cite{giudice}. As already pointed out, the strongest constraint is the one on the stop mass. If we want to avoid big corrections to the soft mass parameters, we have to assume $m_{\tilde{t}} \lesssim  300 $ GeV. Furthermore, if we look at the coupled RG equations for the stop and gluino, to avoid accidental cancellations we have to take $m_{\tilde{g}} \lesssim 2 \div 2.5\mbox{ } m_{\tilde{t}}$.

\begin{figure}[h]
		\begin{center}
	\includegraphics[width=6.5cm]{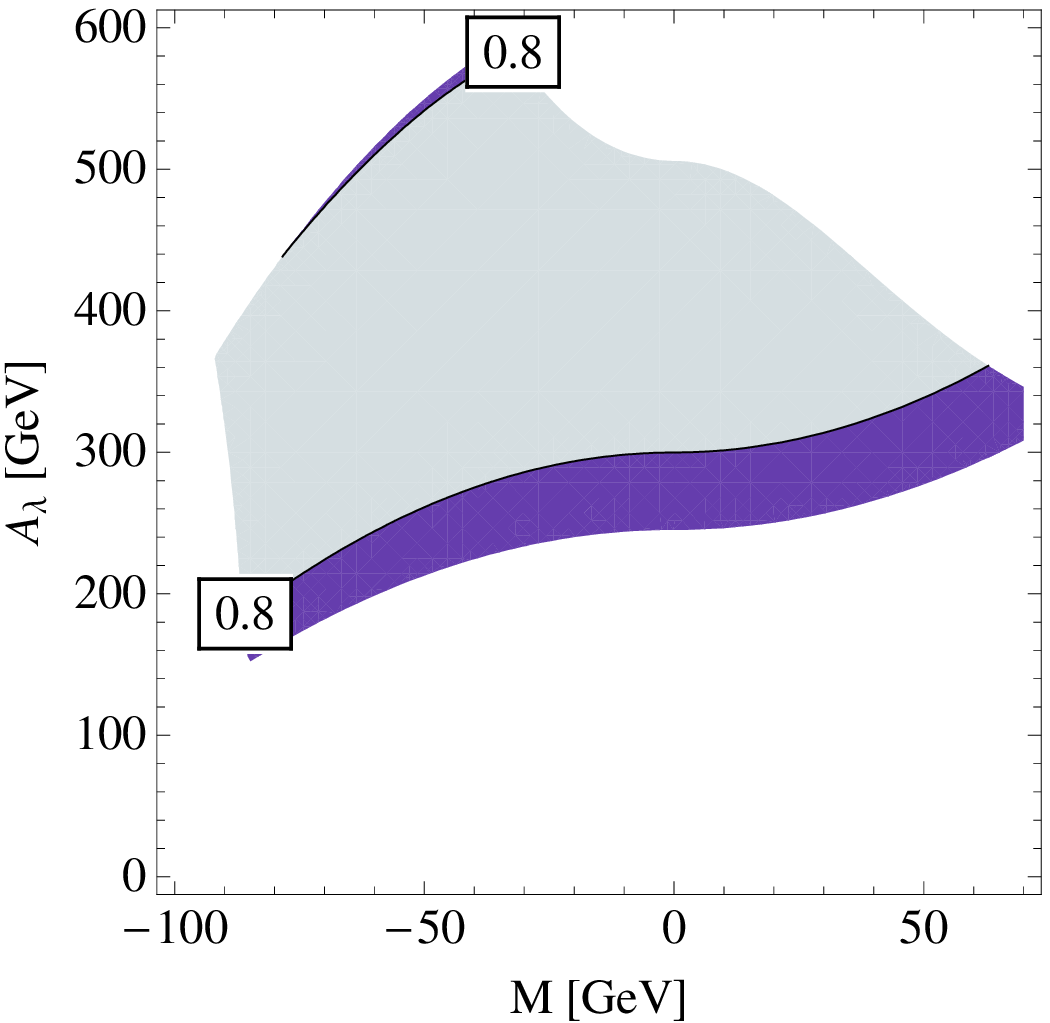}				\includegraphics[width=6.5cm]{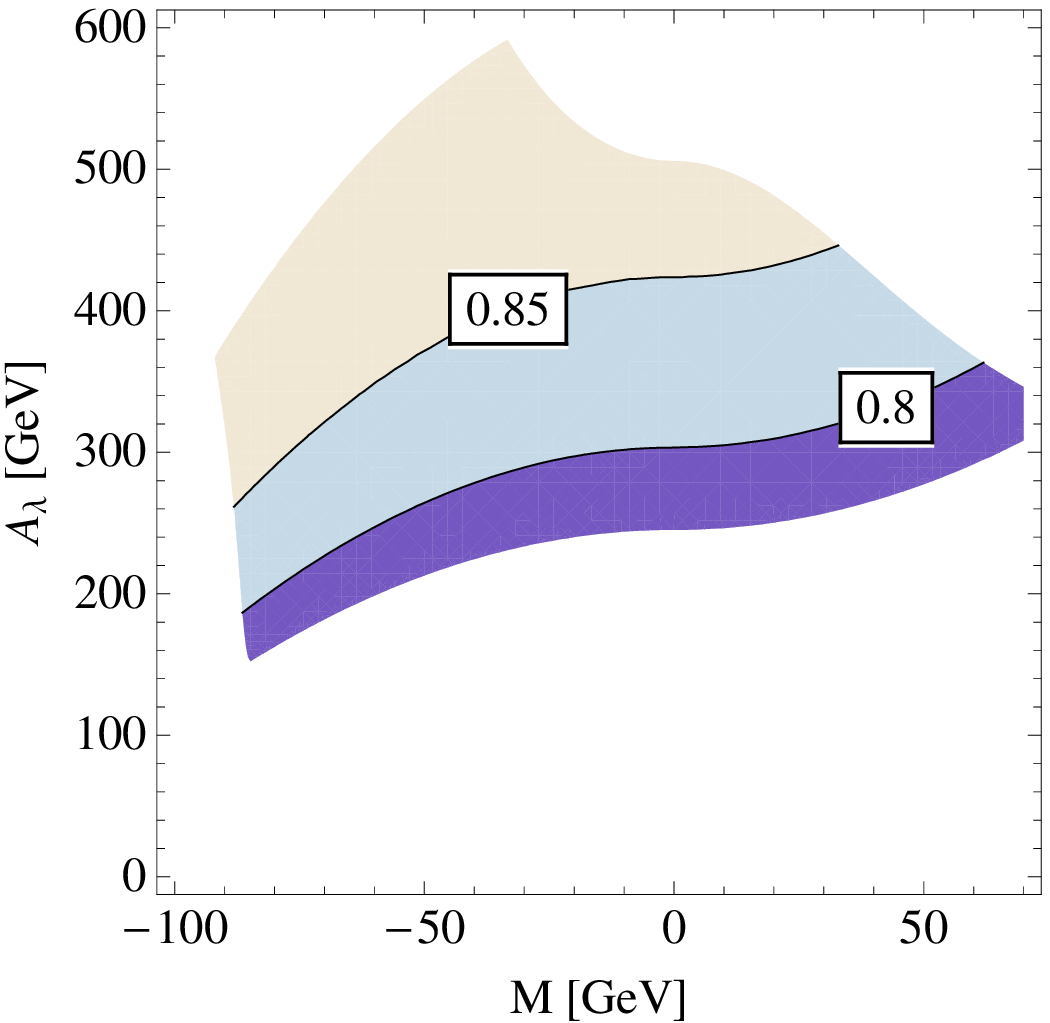}	
			\end{center}
	\caption{Branching ratio of $\tilde{t}_1\rightarrow b \chi^{+}$ for $\tilde{t}_1 \sim \tilde{t}_R$ (left) and $\tilde{t}_1 \sim (1/\sqrt{2})(\tilde{t}_L+\tilde{t}_R)$ (right), as a function of $M = Sgn(m_S^2) \sqrt{|m_S^2|}$  and $A_{\lambda}$ ($A_k = 100$ GeV).}
	\label{stopbr}
\end{figure}

If this interesting limit is realized, the most promising channel for SUSY discovery is the gluino pair production. 
For a gluino mass in the  range $400 \div 600$ GeV the production cross section at the LHC varies between 200 and 10 pb \cite{gluinoxsec}. The produced gluinos then decay with 100\% BR in top-stop pairs, since the stop is the only light squark . The stop decay pattern depends on the details of the stop mixing matrix. As it is shown in Fig. \ref{stopbr} , if the lightest stop is mostly $\tilde{t}_R$ or maximally mixed, it will decay mostly in $b \chi^{+}$, with the chargino further decaying as $\chi^{+} \rightarrow W/W^* \chi_1$. Otherwise, if the $\tilde{t}_1$ is mostly $\tilde{t}_L$ it will decay in $t\chi_1$ with $\sim 100 \%$ branching ratio. The final state thus consists in four $b$-jets, missing energy and some combination of jets and leptons originating by the intermediate state W bosons. The main background sources for this process are $t \bar{t}$, $W/Z$+jets and $bb$+jets. In order to reduce the background cross-sections the particular structure of the final state can be exploited, by imposing cuts on the appropriate variables, such as the number of $b$-tagged jets, $\slashed{E_T}$ and $M_{eff}$. These selection cuts should allow to extract signal from background with a luminosity of the order of 1 fb$^{-1}$ (this preliminary conclusion is supported by the study of similar decay chains  in \cite{gluino1,gluino2}, where however a different mass range was considered). A careful analysis of this process would be of much interest, in order to assess on firm grounds the SUSY discovery potential for this model.

Obviously, a crucial experimental issue is the search for the light scalars, $s_1$ and $s_2$. As was already pointed out in Sec. \ref{coupl}, $s_2$ has in a wide regions of parameters the dominant decay $s_2 \rightarrow a_1 a_1 \rightarrow b \bar{b} b \bar{b}$. This channel is likely to require a high luminosity, even if the gluon fusion cross section is unaffected by the mixing, because of the overwhelming SM backgrounds. The discovery potential for this mode, however, could be enhanced if the associated $W/Z h$ production is considered, as it is pointed out in \cite{Cheung:2007sva}. Other channels worth to be considered, as usual for light scalars, are
$s_2 \rightarrow ZZ^*/WW^*$ and the loop induced decay $s_2 \rightarrow \gamma \gamma$. Even with a small branching ratio, these channels lead to very clean experimental signatures.
The observation of the light scalar $s_1$ seems to be a challenging task, because due to the suppression of the $g_{s_1 VV}$ and $g_{s_1 t \bar{t}}$ couplings (see Fig. \ref{h1vv}) there is a sizable reduction of the production cross sections as well as the decay rates  in the channels  mentioned above. A  detailed study is  needed to assess the discovery potential for both the $s_1$ and $s_2$ particles.

\begin{figure}[h]
		\begin{center}
	\includegraphics[width=9cm]{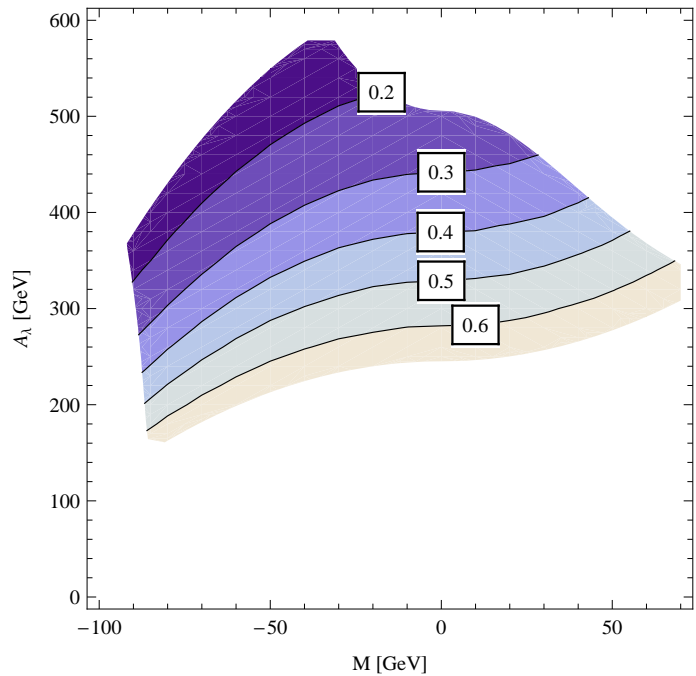}
		\end{center}
		\caption{Cross section for direct production of a chargino-neutralino pair $\sigma(pp \rightarrow \chi_2 \chi^+)$ [pb], as a function of $M = Sgn(m_S^2) \sqrt{|m_S^2|}$  and $A_{\lambda}$ ($A_k = 100$ GeV).}
	\label{xsecneut}
\end{figure}

Finally, another interesting process is the direct neutralino-chargino production $pp \rightarrow \chi_2 \chi^{\pm}$.
The expected production rate for this process is reported in Fig. \ref{xsecneut}.
Without entering into details, it can be roughly estimated that this process would require a luminosity of at least $10$ fb$^{-1}$ in order to be detected in the trilepton channel.

\section{Conclusions}
In this work we studied the generic NMSSM with an R-symmetry invariant superpotential, in the regime where the $\lambda S H_1 H_2$ coupling is taken at the limit of perturbative unification. Under the assumption of extra matter at an intermediate scale, the $\lambda$ value at the weak scale can be as big as $\lambda \approx 0.7 \div 0.8$, allowing to raise the Higgs mass without the need of big radiative corrections. We insist in fact on naturalness, and we thus consider a stop of moderate mass, $m_{\tilde{t}} \lesssim 300$ GeV. 

The R-symmetry in the superpotential forbids explicit mass terms, solving the $\mu$-problem, but allows for the interaction $k S^3$ besides the $\lambda$ term.
We consider the soft supersymmetry-breaking potential that breaks the $R$-symmetry in a general way, including trilinear terms associated to both the $\lambda$ and $k$ supersymmetric couplings. Requiring to have the Higgs boson mass in its maximal range, $115 \div 125$ GeV, we are left with only three effective free parameters: $A_{\lambda}$, $A_k$ and $m_S^2$.

The Higgs boson spectrum contains two light scalars,  $s_1$ and $s_2$, one above and one below the critical value of 115 GeV, a light pseudoscalar $a_1$, and three  heavy states (the scalar $s_3$, the pseudoscalar $a_2$ and two charged particles $H^{\pm}$). The two light scalars, $s_1$ and $s_2$, share the SM coupling to vector bosons pairs $g_{hVV}$. 
Both $s_1$ and $s_2$ prefer to decay in pseudoscalar pairs $a_1 a_1$, whenever this channel is kinematically open.  In the neutralino-chargino sector we assume heavy gaugino masses. In this limit the light states, three neutralinos and one chargino, have only Higgsino components.  For the remaining supersymmetric particles,  we consider the interesting limit where all of the spectrum is heavy except for the states that have to be kept light for naturalness reasons. We thus assume, as already pointed out, the lightest stop to have a moderate mass  ($m_{\tilde{t}_1} \lesssim 300$ GeV) and a gluino not exceeding twice of the $\tilde{t}_1$ mass, while all the remaining squarks and leptons are in the $1 \div 2$ TeV range.

We delimit the allowed region of parameter space where all the LEP2 bounds are respected, insisting on the request to keep the fine-tuning moderate.
The most stringent experimental bound is the one on the light scalar $s_1$.
However, since $s_1$ has a suppressed coupling to vector pairs, there are big regions in parameter space where it would not have been detected at LEP2 even if it is lighter than 115 GeV. On the other hand the naturalness analysis shows that wide regions in parameter space exist where the fine-tuning can be kept below the ten percent level, for a relatively low supersymmetry breaking scale of some hundreds of TeV. For $A_k$ in the range $50 \lesssim A_k \lesssim 150$ GeV, there are sizable regions in the $m_S^2 - A_{\lambda}$ plane where the experimental bounds are respected consistently with a moderate fine-tuning.

We also consider the contribution to the Electroweak Precision parameters S and T from the extended Higgs-Higgsino sector of the model, since $\lambda \approx 0.8$ could introduce a sizable effect. We find that these contributions are perfectly consistent with the experimental data.

The LHC phenomenology of the model depends crucially on the lightest neutralino properties. Naturalness suggests a low supersymmetry breaking scale $\sqrt{F}$, so that  the gravitino is the LSP, while the $\chi_1$ is the next-to-LSP. The width of the $\chi_1 \rightarrow a_1 \tilde{G}$ decay determines the lifetime and the decay length of the NLSP. For $\sqrt{F}$ greater than a few hundreds TeV, the $\chi_1$ will decay outside the detector, and can be thus considered like an effective LSP in the experimental analysis.
In this case we expect SUSY discovery in the gluino pair production process with $\sim 1$ fb$^{-1}$ of luminosity or less, depending on the gluino mass.
Other interesting processes at higher luminosity are the production of the light scalars and Higgsinos. A detailed analysis of these signatures  would be of great interest to assess the LHC discovery potential for this model.

\section*{Acknowledgments}
I would like to thank Riccardo Barbieri for inspiring this study, and as well for carefully reading the manuscript and for useful discussions. I also thank Paolo Azzurri, Brando Bellazzini and Roberto Franceschini for useful discussions.

\appendix

\section{One loop contributions to S and T}
\label{stloop}
We report here for the reader's convenience the expressions \cite{Barbieri:1983wy,lsusy} of the one loop function that appear in the contributions to S and T from new particles.

\noindent For a boson loop with internal masses $m_1, m_2$, the functions $A, F, G$ appearing in Eqs. (\ref{thiggs}, \ref{sthiggs}) are given by:
\begin{align}
   A(m_{1},m_{2}) = 
  & \frac{1}{32\pi^{2}  \alpha_{\rm em} v^{2}} \biggl[ \frac{m_{1}^{2}+m_{2}^{2}}{2}
    - \frac{m_{1}^{2}m_{2}^{2}}{m_{1}^{2}-m_{2}^{2}}
      \ln\frac{m_{1}^{2}}{m_{2}^{2}} \biggr] ,
\\
  F(m_{1},m_{2}) = 
  & \frac{1}{24\pi} \biggl[ -\ln\frac{\Lambda^{4}}{m_{1}^{2}m_{2}^{2}}
    + \frac{4m_{1}^{2}m_{2}^{2}}{(m_{1}^{2}-m_{2}^{2})^{2}}
\nonumber\\
  & + \frac{m_{1}^{6}+m_{2}^{6}-3m_{1}^{2}m_{2}^{2}(m_{1}^{2}+m_{2}^{2})}
      {(m_{1}^{2}-m_{2}^{2})^{3}} \ln\frac{m_{1}^{2}}{m_{2}^{2}} \biggr] ,
\\
G(m_{1},m_{2}) = 
& \frac{1}{2\pi}
   \left[ \frac{2m_{1}^{2}m_{2}^{2}}{(m_{1}^{2}-m_{2}^{2})^{3}}
     \ln\frac{m_{1}^{2}}{m_{2}^{2}}
   - \frac{m_{1}^{2}+m_{2}^{2}}{(m_{1}^{2}-m_{2}^{2})^{2}} \right] .
\end{align}

\noindent For a fermion loop with internal masses $m_1, m_2$, the functions $\tilde{A}, \tilde{F}$ appearing in Eqs. (\ref{thiggsino}, \ref{sthiggsino}) are given by:
\begin{align}
\tilde{A}(m_{1},m_{2}) =
  & \frac{1}{32\pi^{2} \alpha_{\rm em} v^{2}} \biggl[ (m_{1}-m_{2})^{2}\ln\frac{\Lambda^{4}}
      {m_{1}^{2}m_{2}^{2}} - 2m_{1}m_{2}
\nonumber\\
  & + \frac{2m_{1}m_{2}(m_{1}^{2}+m_{2}^{2})-m_{1}^{4}-m_{2}^{4}}
      {m_{1}^{2}-m_{2}^{2}} \ln\frac{m_{1}^{2}}{m_{2}^{2}} \biggr] ,
\\
  \tilde{F}(m_{1},m_{2}) =
  & \frac{1}{6\pi} \biggl[ -\ln\frac{\Lambda^{4}}{m_{1}^{2}m_{2}^{2}}
    - \frac{m_{1}m_{2}(3m_{1}^{2}-4m_{1}m_{2}+3m_{2}^{2})}
      {(m_{1}^{2}-m_{2}^{2})^{2}}
\nonumber\\
  & + \frac{m_{1}^{6}+m_{2}^{6}-3m_{1}^{2}m_{2}^{2}(m_{1}^{2}+m_{2}^{2})
      +6m_{1}^{3}m_{2}^{3}}{(m_{1}^{2}-m_{2}^{2})^{3}} 
      \ln\frac{m_{1}^{2}}{m_{2}^{2}} \biggr] .
\end{align}

\end{document}